\font\fontA=cmss17
\def\be{\begin{equation}}
\def\ee{\end{equation}}
\def\bea{\begin{eqnarray}}
\def\eea{\end{eqnarray}}

\newcommand{\btheta}{\mbox{\boldmath $\theta$}}
\newcommand{\brho}{\mbox{\boldmath $\rho$}}
\def\ni{\hfil\break\noindent}
\documentstyle[12pt,a4]{article}
\begin{document}
\thispagestyle{empty}
\rightline{\large\baselineskip16pt\rm\vbox to20pt{\hbox{OCHA-PP-53}
%\hbox{\today}
\vss}}
\vskip15mm
\begin{center}
{\fontA
Spontaneous dissipation from generalized radiative corrections
}
\end{center}
\begin{center}
{\sc Masahiro~MORIKAWA } \\[2mm]
{\it Department of Physics, Ochanomizu university \\[2mm]
1-1, Otsuka 2, Bunkyo-ku, Tokyo 112 JAPAN  \\
e-mail: hiro@phys.ocha.ac.jp}
\vskip1truecm
({\it received ~~~~~~~~~~~~~~~ 1995})
\end{center}
\begin{abstract}
We derive dissipative effective Hamiltonian for the unstable Lee model without
any ad hoc coarse graining procedure.  Generalized radiative corrections,
utilizing the in-in formalism of quantum field theory, automatically yield
irreversibility as well as the decay of quantum coherence.  Especially we do
not need to extend the ordinary Hilbert space for describing the intrinsically
dissipative system if we use the generalized in-in formalism of quantum field
theory.
\ni
pacs numbers: 	%
05.70.Ln,~~ 	% non equilibrium, irreversible
03.70.+k ,~~	% field theory
82.20.Mj,~~	% non equilibrium kinetics
11.10.Wx	% thermal field theory

\end{abstract}
\vskip1cm
%\multicols{2}
%
%
%
%
%
%
%%%  section 1 %%%%%%%%%%%%%%%%%%%%%%%%%%%%%%%
\section{Introduction}
%%%%%%%%%%%%%%%%%%%%%%%%%%%%%%%%%%%%%%%
Understanding the irreversibility in the macroscopic world is one of the most
attractive issues in physics.  Especially a consistent derivation of the
irreversible dynamics from much fundamental microscopic law of physics would be
the central issue.  Our ill fortune is that the most microscopic physics are
strictly reversible and a simple application of them never yield
irreversibility.
\par
A popular approach to obtain irreversibility will be to consider an open
system\cite{li86}:  We decompose the total closed system into a relevant system
and the remaining environmental degrees of freedom.  Then by {\it coarse
graining} the environmental degrees of freedom with appropriate initial
conditions (projection), we obtain effective dynamics for the relevant system.
The irreversibility stems from the information loss of the system into the
environment.  Although this pragmatic procedure is widely used in the
literature, qualitative dissipative nature in general depends on how we set the
separation of the total system and on the coarse graining procedure.  Surely
this is unfavorable nature of the theory; the irreversibility is an intrinsic
nature of the system and should not be affected by the method of description.
In this paper, we would like to demonstrate that for certain systems, it is
possible to derive intrinsic irreversibility without specifying the separation
and the coarse graining methods

\par
It is obvious that not all the systems show irreversibility.  Then what is the
essential difference between reversible and irreversible systems?  According to
our experience, a system should, at least, be {\it unstable} for it to show
irreversibility.  It is manifest that a stable state cannot change further and
shows no irreversibility.  On the other hand the unstable system cannot persist
on the initial state and eventually decays into much stable state if any.
However instability will not be the sufficient condition for the
irreversibility.  If the stable-unstable transition is simple and the system
has finite recursion time, then the system cannot be irreversible.  In order to
obtain the infinite recursion time, we need {\it infinite number of degrees of
freedom or chaos}\cite{kubotani95}.  Another necessary condition for
irreversibility will be a {\it natural averaging} procedure whatever it is
implicit or explicit.  One such averaging procedure will be the radiative
corrections in quantum theory.  Esp

a) instability
b) infinite degrees of freedom and
c) natural averaging,
are sufficient conditions for the irreversibility, we try to use a simple model
which satisfies all the above conditions.  It is the unstable Lee
model\cite{lee54}.  This simple model of quantum field theory is exactly solved
and was used in the argument of renormalizations.  In this article, we use this
unstable Lee model and demonstrate a possible origin of the intrinsic
dissipative dynamics without any ad hoc coarse graining procedure.
\par
The {\it instability} of the system is considered to be an essential ingredient
for the emergence of intrinsic dissipativity.  However, an unstable system can
be described by hermitian Hamiltonian which includes no dissipativity nor
irreversibility at least in appearance.  There is a long history of the study
on unstable states and its decay in quantum mechanics, in particle physics
\cite{nakanishi58} \cite{sudarshan78}  and in statistical mechanics
\cite{petrosky91} \cite{antoniou93}.  They faced with the complex eigenvalue
for the Hamiltonian, and therefore extended the Hilbert space so that to
maintain the hermiticity of the Hamiltonian.  In this procedure they abandon
the usual Hilbert space (a space of square integrable functions) and introduced
a rigged Hilbert space (a space of distributions).  According to this method, a
pair of dual spaces $\Psi_+^{\dagger}$ and $\Psi_-^{\dagger}$ \cite{antoniou93}
is necessary.  They are dual of the space of boundary functions which are
analytic in the lower (upper

\par
We will not use this extended Hilbert space approach in the present article in
order to describe the intrinsic dissipative nature of the unstable system.
However, we use the generalized in-in formalism of quantum field theory
\cite{schwinger61} \cite{keldysh64} \cite{chou85}, in which dissipative and
irreversible properties are consistently incorporated through radiative
corrections\cite{morikawa86} \cite{morikawa95}.  In the ordinary quantum field
theory (in-out formalism), ultra-violet divergence in the quantum system of
infinite degrees of freedom necessitates the renormalization of the parameters
in the Hamiltonian through the process of radiative corrections.  This
radiative correction is a special kind of averaging, which does not directly
yield irreversibility in the stable theory.  If the system is unstable, this
process yields complex poles in the retarded propagator.  The location of a
pole is interpreted as the decay strength or the inverse of the lifetime.  This
dissipativity was not present in

\par
In this paper, we first review the unstable Lee model in the usual treatment in
the next section \S 2.  Then we study the same model in the in-in formalism of
quantum field theory in \S 3 and derive a Langevin equation for the fields.  We
further derive the effective Hamiltonian which describes irreversible dynamics
in  \S 4 and show that the linear entropy automatically increases.  The last
section \S 5 is for discussions and summary of our work.
%
%
%
%
%
%
%
%%%%   section 2   %%%%%%%%%%%%%%%%%%%%%%%%%%%%%
\section{Unstable Lee model}
%%%%%%%%%%%%%%%%%%%%%%%%%%%%%%%%%%%%%%%
\par
Let us begin our argument from the unstable Lee model in the usual treatment.
The system is composed from two kinds of non-relativistic fermions ${\bf N}$
and ${\bf V}$ and one boson field ${\btheta}$.

Hamiltonian of the Lee model is given by
\bea
{\bf H}&=&{\bf H}_0 + {\bf H}_{int} \nonumber \\
{\bf H}_0&=&m_V^0 \int {d\vec p \over (2\pi)^3}
{\bf V}_{\vec p}^{\ast} {\bf V}_{\vec p}
+ m_N \int {d\vec p \over (2\pi)^3}{\bf N}_{\vec p}^{\ast} {\bf N}_{\vec p}
+\int {d\vec k \over (2\pi)^3} \omega_{\vec k}\btheta_{\vec k}^{\ast}
\btheta_{\vec k}\nonumber \\
{\bf H}_{\rm int}&=&\lambda_0 \int {d\vec k \over (2\pi)^3 \sqrt{2 \omega_{\vec
k}}} \int {d\vec p \over (2\pi)^3}
{f(\omega_{\vec k})} ({\bf V}_{\vec p}^{\ast} {\bf N}_{\vec p -\vec
k}\btheta_{\vec k} + {\bf h.c.}),
\label{bhamiltonian}
\eea
where
$\omega_{\vec k}=\sqrt{\vec k^2 +\mu^2}$.
The states
$| N_{\vec p}\rangle \equiv {\bf N}_{\vec p}^{\ast}|0\rangle$ and
$| a_{\vec k}\rangle \equiv {\bf a}_{\vec k}^{\ast}|0\rangle$ are the
eigenstates of the total Hamiltonian, however the state
$| V_{\vec p}\rangle \equiv {\bf V}_{\vec p}^{\ast}|0\rangle$ is not.
\par
In order to construct the eigenstate $| {\tilde V}_{\vec p}\rangle$, we have to
superpose the states:
$| V_{\vec p}\rangle$ and
$| a_{\vec k} N_{\vec p-\vec k}\rangle$ as
\be
| {\tilde V}_{\vec p}\rangle=\sqrt{Z_V}
\left( | V_{\vec p}\rangle + \int {d\vec k \over (2\pi)^3} g_{\vec k} | a_{\vec
k} N_{\vec p-\vec k}\rangle \right).
\label{vstate}
\ee
Then the weight $g_{\vec k}$ in this form is determined from the eigenstate
equation
\be
{\bf H}| {\tilde V}_{\vec p}\rangle = m_V | {\tilde V}_{\vec p}\rangle,
\label{eigensteq}
\ee
as
\be
g_{\vec k} ={1 \over m_V-m_N-\omega_{\vec k}}
{ \lambda_0 f(\omega_{\vec k}) \over \sqrt{2\omega_{\vec k}} }.
\label{weight}
\ee
The determination of the eigenvalue $m_V$ is equivalent to looking for a root
of the retarded propagator:
\be
G_R(E)=\left[ E- m_V^0+ i \epsilon
+\int {d\vec k \over (2 \pi)^3}  {1 \over m_N+\omega_{\vec k}-E-i \epsilon}
{ \lambda_0^2  f^2(\omega_{\vec k}) \over 2\omega_{\vec k} }\right]^{-1}.
\label{invret}
\ee
A real root is found for a stable $V$- particle.  However if unstable, the root
becomes complex: $E=m_V- i \gamma/2$.
Usually $m_V$ is interpreted as the observable mass and $1/\gamma$ as the
life-time of the physical $V$ particle\cite{gamow28}.  However, the state  $|
{\tilde V}_{\vec p}\rangle$ turns out to have zero norm $\langle {\tilde
V}_{\vec p}| {\tilde V}_{\vec p}\rangle=0$ simply because the Hamiltonian is a
hermitian operator.  Nakanishi and others \cite{nakanishi58} \cite{sudarshan78}
expressed the eigenstate corresponding to this eigenvalue by introducing the
notion of {\it complex distribution}.  On the other hand in papers
\cite{petrosky91} \cite{antoniou93}, they extended the ordinary Hilbert space
introducing  the notion of the {\it rigged Hilbert space}.
%
%
%
%
%
%
%%%  section 3      %%%%%%%%%%%%%%%%%%%%%%%%%%%%%
\section{Unstable Lee model in the in-in quantum field theory}
%%%%%%%%%%%%%%%%%%%%%%%%%%%%%%%%%%%%%%%
\par
We will take a conservative approach.  Instead of extending the ordinary
Hilbert space,  we express the evolution of the unstable particle in the in-in
formalism of quantum field theory \cite{schwinger61} \cite{keldysh64}
\cite{chou85}, which is the most appropriate formalism for describing the
unstable quantum system\cite{morikawa86}.  This is just a simple extension of
the ordinary quantum field theory with the doubled time-contour of integration.
 In this formalism, it is possible to express the statistical dissipation and
fluctuations consistently with quantum field theory.  Leaving the detail of
this formalism for the other paper \cite{morikawa95}, we briefly explain this
formalism here.
\par
The time contour of integration in the in-in quantum field theory is
generalized to run from $-\infty$ to $+\infty$ and then back to $-\infty$
again.
All the arguments of fields ${\bf \Phi}(x)$ are doubled according to this new
time contour.
Moreover, the Hamiltonian ${\bf H}[{\bf \Phi}]$ of the system is generalized to
$\hat {\bf H}[{\bf \Phi}^{\pm}]={\bf H}[{\bf \Phi}^+]-{\bf H}[{\bf \Phi}^-]$
where $X^+(x)$ and $X^-(x)$ mean the field quantity $X(x)$ restricted on the
forward and backward time branches, respectively.
In the same manner, the Lagrangian density ${\cal L}$ is generalized to
$\hat{\cal L}={\cal L}[{\bf \Phi}^+]-{\cal L}[{\bf \Phi}^-]$.  Because the
standard Pauli equation for the density matrix
$i \partial \brho(t)/\partial t=[{\bf H}, \brho(t)]$ is expressed in the
coordinate representation ($\langle \Phi_+|\brho(t)|
\Phi_-\rangle=\rho[\Phi^{\pm},t]$) as
\be
i {\partial \rho[\Phi^{\pm},t] \over \partial
t}=(H[\Phi^+]-H[\Phi^-])\rho[\Phi^{\pm},t],
\label{pauli}
\ee
this generalized Hamiltonian, in the coordinate representation, $\hat H$ is
thought to be the time translation operator for the density matrix.
\par
The partition function is defined in the usual way except that the
time-integration contour is doubled.
\bea
\hat Z[J]&\equiv& {\rm Tr}[ T_C({\rm exp}[i\ \int_C d^4 x { J(x) {\bf \Phi}(x)
}] )\rho]
\nonumber \\
&=&{\rm Tr}[ T_+({\rm exp}[i\ \int d^4 x{ J_+(x) {\bf \Phi}_+(x) }])T_-({\rm
exp}[-i\ \int d^4 x{ J_-(x) {\bf \Phi}_-(x) })])\rho]
\label{pfunction}
\eea
where the suffix $C$ in the integral means that the time integration contour is
generalized so that it runs from minus infinity to plus infinity and then back
to the minus infinity again.   The symbol $\rho$ is the initial density matrix.
 The symbol ${\bf \Phi}(x)$ represents all the quantum fields in Heisenberg
picture.  Generalized effective action $\hat\Gamma[\Phi]$ is defined simply as
the Legendre transformation of the above partition function $\hat Z[J]$.
Perturbation method using generalized propagators is available for calculating
various quantities.
\par
Back to the unstable Lee model, we calculate the generalized effective action.
Because there is only one loop correction for the $V$- particle propagator and
no correction for the $N$- and boson- particle propagators, the calculation is
exactly done and the result is
\bea
\hat\Gamma&=&S_N[N_+]-S_N[N_-]+S_{\theta}[\theta_+]-S_{\theta}[\theta_-]
+S_{\rm int}[N_+, V_+,\theta_+]-S_{\rm int}[N_-, V_-,\theta_-] \nonumber\\
&+&\int d^4x \int d^4x'
\left( \Phi_V^+,  \Phi_V^-\right)^{\ast}_x
\left( {\matrix{D-iB&i(B-A)\cr
i(B+A)&-D+iB\cr
}} \right)_{x,x'}
\left( {\matrix{
\Phi_V^+\cr
\Phi_V^-}} \right)_{x'}
\label{effaction}
\eea
where all $N$ and $V$ variables are  Grassmann valued fields.
In the above, the first line represents the bare actions corresponding to
Eq(\ref{bhamiltonian}) but with renormalized coupling constant $\lambda$
instead of $\lambda_0$.
The last line is the radiatively corrected $V$-particle part and is further
rewritten as
\be
\int d^4x \int d^4x'
\left( \Phi_V^{\Delta},  \Phi_V^C\right)^{\ast}_x
\left( {\matrix{iB&D+iA\cr
D-iA&0\cr
}} \right)_{x,x'}
\left( {\matrix{
\Phi_V^{\Delta}\cr
\Phi_V^C}} \right)_{x'},
\label{veffaction}
\ee
where
$\Phi_{\Delta}=\Phi_+-\Phi_-$ and $\Phi_C=(\Phi_++\Phi_-)/2$.
Kernels A, B, and D are induced from radiative corrections and are exactly
calculated.   Their Fourier transforms are given by
\bea
D(E)&=&E-m_V^0+\int {d \vec k \over (2\pi)^3}
{\lambda_0^2 f^2(\omega_{\vec k}) \over 2 \omega_{\vec k}}
{{\cal P} \over m_N+\omega_{\vec k}- E}, \nonumber\\
B(E)&=&Z_V^{-1}\theta(E-m_N-\mu) \sqrt{(E-m_N)^2-\mu^2} \lambda^2 f(E-m_N)/(4
\pi), \nonumber\\
A(E)&=&{\rm sign}(E)B(E),
\label{kernels}
\eea
in the momentum representation but the three momentum is suppressed.
$D(E)$ part yields the infinite mass correction and the wave function
renormalization as
\be
D(E)=(1+C_1)E-(m_V^0-C_0 +C_1 m_V)=Z_V^{-1} E -m_V,
\label{dpart}
\ee
where
\be
C_0= \int {d\vec k \over (2 \pi)^3}
{ \lambda_0^2  f^2(\omega_{\vec k}) \over 2\omega_{\vec k} }
{{\cal P} \over m_N+\omega_{\vec k}-m_V}, ~~
C_1=\int {d\vec k \over (2 \pi)^3}
{ \lambda_0^2  f^2(\omega_{\vec k}) \over 2\omega_{\vec k} }
{{\cal P} \over (m_N+\omega_{\vec k}-m_V)^2}.
\label{coeff}
\ee
The renormalized coupling constant $\lambda$ and the wave function
renormalization $Z_V$ appeared in Eq.(\ref{kernels}) are defined as
\be
Z_V^{-1}=1+C_1, ~~~ \lambda^2=Z_V \lambda_0^2.
\label{renormalization}
\ee
The $B(E)$ part comes from the quantum cross correlation between the forward
time branch and the backward time branch.  This term was absent in the usual
in-out formalism and is new in the in-in formalism.  The $A(E)$ part is also
specific to the in-in formalism and it breaks time reversal symmetry because it
is odd in the argument $E$.  The time irreversible term $A(E)$ appears simply
because we are considering specific boundary condition:  We have taken {\it
initial} density matrix $\rho$ in Eq.(\ref{pfunction}).  If we took the {\it
final} density matrix there, the signature of the $A(E)$ term would be
reversed.
\par
The above kernels are in general non-local.  Here we take the local
approximation (setting $E \rightarrow m_V$) with the limit $m_V \gg m_N, \mu$
just for simplicity.\footnote{%
This non-locality means that the retarded effect has finite time scale and it
turns out later that the natural noise associated with the system is colored. }
Then the kernels $A$ and $B$ become
\be
A(E) \approx {\lambda^2 f(m_V) \over 4\pi} E, ~~~~
B(E) \approx {\lambda^2 f(m_V)  \over 4\pi}m_V,
\label{localapp}
\ee
where we have set $Z_V=1$.
Note that the effective action becomes complex reflecting the fact that the
system is unstable.\footnote{%
Because $A(E)$ is odd ($A(-E)=-A(E)$) and $B(E)$ is even ($B(-E)=B(E)$), their
Fourier transforms $A(t)$ and $B(t)$ are pure imaginary and real, respectively.
}
The pure imaginary term, which is proportional to $B(t)$, is however symmetric
for the exchange of the variables $\Phi_V^+ \leftrightarrow \Phi_V^-$; all
other terms are anti-symmetric.  Therefore if we define the hermitian conjugate
including the operation of the exchange of $\Phi_V^{\pm}$, the generalized
effective action $\hat\Gamma$ becomes hermitian.  In fact, this hermiticity is
explicitly realized later in the effective Hamiltonian.
\par
Now we derive the generalized equations of motion for the fields.  Because only
the $V$- filed shows irreversibility and dissipativity, we concentrate on this
field and suppress the suffix $V$ for the moment.  The above effective action
can be re-expressed in a cute form which manifestly represents dissipativity if
we introduce auxiliary fields $\xi(t)$ and $\xi^{\ast}(t)$ which are also
Grassmann valued fields.  If we decompose the effective action as
$ \Gamma ={\rm Re} \Gamma +i{\rm Im} \Gamma $,
then the imaginary part is even in the variable $\Phi _\Delta (x) $:
\be
{\rm Im} \hat\Gamma [\Phi _c,\Phi _\Delta,\Phi^{\ast} _c,\Phi^{\ast} _\Delta ]
=\int\!\!\!\int {\Phi^{\ast}_\Delta(x)B(x-y)\Phi _\Delta (y)}.
\label{imeaction}
\ee
We can rewrite this expression by introducing auxiliary fields $\xi(x)$ and
$\xi^{\ast}(x)$ which are Grassmann valued fields
\be
{\rm exp}[i\hat\Gamma [\Phi, \Phi^{\ast} ]]
=\int {[d\xi] [ d\xi^{\ast} ]}P[\xi, \xi^{\ast} ]
{\rm exp}[i {\rm Re} \Gamma +\int ({i\Phi^{\ast} _\Delta \xi-i
\xi^{\ast}\Phi_\Delta})]
\label{decomposition}
\ee
where,
\be
P[\xi, \xi^{\ast} ]= ({\rm det}B){\rm exp}[\int \!\! \int {\xi^{\ast} B^{-1}\xi
}]
\label{weightp}
\ee
is a normalizable positive kernel for the fields $\xi(x)$ and $\xi^{\ast}(x)$.
Note that this weight function is purely Gaussian.
It means that we may be able to interpret $P[\xi, \xi^{\ast}]$ as a statistical
weight for the random fields $\xi(x), \xi^{\ast}(x)$(stochastic part).
Therefore it is possible to interpret Eq.(\ref{decomposition}) that the total
effective action $\Gamma$ is a statistical average of the individual effective
actions ${\rm Re} \Gamma -\int {\xi^{\ast} \Phi _\Delta }+\int {\Phi^{\ast}
_\Delta \xi }$.
Application of the variational principle on this individual effective action
yields an equation of motion for
$\Phi_C(x)$ as
\bea
0&=& \left({{\delta  {\rm Re} \Gamma -\int {\xi^{\ast} \Phi _\Delta }+\int
{\Phi^{\ast} _\Delta \xi }}
	\over {\delta  \Phi^{\ast}_{\Delta} (x)}} \right)_{\Phi_{\Delta}=0} \nonumber
\\
&=& \left( (i-\gamma)\partial_t-m_V+{\nabla^2 \over 2m_V}\right)\Phi_C + V'+
\xi,
\label{langevin}
\eea
where we have used the local approximation and $V'$ is the interaction term in
the total Hamiltonian $H_{\rm int}$.  We have set $J=0$, which means there is
no external source.  The symbol $\gamma$ is $\lambda_0^2 f(m_V)/(4\pi)$.  This
is a renormalized Langevin type stochastic differential equation with friction
and random force terms.
According to this equation, the evolution of the field $\Phi_C(x)$ is partially
deterministic and partially stochastic.  The former part is governed by the
action $ {\rm Re} \Gamma [\Phi ]$ which include the damping effect and the
latter part is induced by the random field $\xi(x)$ whose statistical
properties are completely determined by ${\rm Im} \Gamma [\Phi ]$.
Actually if we define the statistical average  as
\be
\left\langle {\cdots} \right\rangle _{\xi,\xi^{\ast}} \equiv \int {[d\xi
][d\xi^{\ast} ]}P[\xi, \xi^{\ast} ]\cdots,
\label{statav}
\ee
then we obtain the correlation function for the random field
\be
\left\langle {\xi^{\ast} (x)\xi (y)} \right\rangle _\xi =B(y-x),
\label{correlation}
\ee
which becomes white noise if we take the local approximation
Eq.(\ref{localapp}).
The same variational principle yields the equations of motion for the other
fields $N$ and $\theta$.  However there appear no new terms which show
dissipativity and irreversibility even after full radiative corrections.
%
%
%
%
%
%
%
%%%%   section 4   %%%%%%%%%%%%%%%%%%%%%%%%%%%%%
\section{Effective Hamiltonian and entropy increase}
%%%%%%%%%%%%%%%%%%%%%%%%%%%%%%%%%%%%%%%
\par
We can express the dissipativity of the system in another form by constructing
the effective Hamiltonian.  If we apply the local approximation, then the
effective action reduces to the local form $\hat \Gamma_V=\int d^4x \hat{\cal
L}_V$, where ${\cal L}_V$ is the generalized effective Lagrangian for the
$V$-particle part:
\bea
\hat {\cal L}_V
&=&i (\Phi^{\Delta\ast}_V \dot\Phi^C_V +\Phi^{C\ast}_V\dot \Phi^{\Delta}_V   )
+\Phi^{\Delta\ast}_V {\nabla^2 \over 2m_V} \Phi^C_V
+\Phi^{C\ast}_V {\nabla^2 \over 2m_V} \Phi^{\Delta\ast}_V
\nonumber\\
&-&\gamma (\Phi^{\Delta\ast}_V \dot\Phi^C_V-\Phi^{C\ast}_V
\dot\Phi^{\Delta\ast}_V) +i \gamma  m_V \Phi^{\Delta\ast}_V  \Phi^{\Delta}_V.
\label{efflagrangian}
\eea
The canonical momenta are defined by
$p_V^{\pm}\equiv \pm \partial\hat{\cal L}/\partial \dot\Phi_V^{\pm}$ or
\be
p_V^{\Delta} \equiv {\partial \hat{\cal L} \over \partial\dot\Phi_V^{\Delta}}
=(i-\gamma) \Phi_V^{C\ast}, ~~
p_V^C \equiv {\partial \hat{\cal L} \over \partial\dot\Phi_V^C}
=(i+\gamma) \Phi_V^{\Delta\ast}.
\label{canmom}
\ee
Then the generalized effective Hamiltonian for the total system becomes
\bea
\hat H&=& \int d^3 x
[p_V^{\Delta}\dot\Phi_V^{\Delta}+p_V^C\dot\Phi_V^C
- \hat {\cal L}]    \nonumber\\
&=&
H[\Phi_N^+,\Phi_{\theta}^+]-H[\Phi_N^-,\Phi_{\theta}^-]
+H_{\rm int}[\Phi^+]-H_{\rm int}[\Phi^-] \nonumber\\
&-&\Phi^{\Delta\ast}_V {\nabla^2 \over 2m_V} \Phi^C_V
-\Phi^{C\ast}_V {\nabla^2 \over 2m_V} \Phi^{\Delta\ast}_V
-i \gamma  m_V \Phi^{\Delta\ast}_V  \Phi^{\Delta}_V.
\label{effhamiltonian}
\eea
where $H[\Phi_N^{\pm},\Phi_{\theta}^{\pm}]$ and $H_{\rm int}[\Phi^{\pm}]$ are,
respectively, the free $N~\theta$- particle part and the interaction part of
the original Hamiltonian Eq.(\ref{bhamiltonian}).  There is no radiative
corrections for these parts except $Z_V$ and $\lambda$.  Note that the
effective Hamiltonian, even after the local approximation, is hermitian in the
sense $H[\Phi^-,\Phi^+]^*=H[\Phi^+, \Phi^-]$.
Therefore if we write down the generalized Pauli equation as
$i {\partial \rho[\Phi^{\pm}] / \partial t}=\hat
H[\Phi^{\pm}]\rho[\Phi^{\pm}]$,
the total probability is conserved (${\rm Tr}\rho={\rm const.}$).
We now demonstrate this in an operator form of the Pauli equation.
\par
We rewrite the above Pauli equation for the density matrix in an operator form.
 Remember that we have been using the representation:
${\bf \Phi}(\vec x)| \Phi\rangle=\Phi(\vec x) | \Phi\rangle$,
$\langle \Phi_+| \brho(t) | \Phi_-\rangle=\rho[\Phi^{\pm},t]$, and so on.
Therefore for example, the operator form of $\Phi_{\Delta}~\rho[\Phi^{\pm},t]$
is
$[{\bf \Phi},  \brho(t)]$.
In the similar way, the operator form of the Pauli equation becomes
\be
i{\partial  \brho(t)\over \partial t}
=[{\bf H}_R,  \brho(t)]
-i \gamma m_V [{\bf \Phi}^{\ast}_V , [ {\bf \Phi}_V,  \brho(t)]],
\label{opform}
\ee
where $H_R$ is the total Hamiltonian Eq.(\ref{bhamiltonian}) with
renormalizations.  It is easy to see the conservation of probability from this
equation:
\be
{\partial \over \partial t}{\rm Tr} \brho(t)=0.
\label{consprob}
\ee
Further we define the linear entropy $S(t) \equiv -{\rm Tr} \brho(t)^2$, which
has values in $[-1,0]$.  This measures how the system possesses coherence or
the amount of classicality; $S(t)=-1$ for a pure quantum state, and the
coherence is much destroyed for larger values of $S(t)$\cite{morikawa90}.
Time evolution of $S(t)$ is governed by the last term on the RHS in
Eq.(\ref{opform}),
\bea
{\partial S(t)\over \partial t}&=&2\gamma m_V {\rm Tr}[ \brho(t), {\bf
\Phi}^{\ast}_V][{\bf \Phi}_V, \brho(t)]
\nonumber\\
&=&2\gamma m_V {\rm Tr}|[{\bf \Phi}_V,  \brho(t)]|^2 > 0.
\label{lentropy}
\eea
Therefore the linear entropy perpetually increases.
This manifestly shows irreversibility of the system.
We emphasize that we did not use any ad hoc averaging method such as partial
trace on the environment; we do not have any environment at all.  Therefore the
irreversibility represented by Eq.(\ref{lentropy}) is intrinsic to the system.
Moreover this result does not rely upon ad hoc approximations such as the
truncation of the sequence of correlation functions\footnote{%
We have used non-relativistic approximations, neglected the recoil of the heavy
$V$-particle, and took local approximations.  These are all our approximations.
}%
; the Lee model is exactly solved.
This decoherence term stems from the quantum interference between the fields on
the forward time branch and those on the backward time branch, and was not
exist in the usual quantum field theory of in-out formalism.
%
%
%
%
%
%%%  section 5 %%%%%%%%%%%%%%%%%%%%%%%%%%%%%%%
\section{Discussions and summary}
%%%%%%%%%%%%%%%%%%%%%%%%%%%%%%%%%%%%%%%
\par
In this article, we have arrived at the dissipative expressions
Eq.(\ref{langevin})Eq.(\ref{effhamiltonian})Eq.(\ref{opform}) for the dynamics
of unstable Lee model.  We would like to emphasize the following points for the
origin of intrinsic irreversibility of the model.
\begin{enumerate}
\item
The starting point of our study has been the bare Hamiltonian
Eq.(\ref{bhamiltonian}), in which no dissipativity is manifest.  However this
bare Hamiltonian itself does not correctly describe the real system; we need to
take into account the radiative corrections and remove the divergences in the
theory.  These radiative corrections and renormalizations are essential for
defining a feasible theory.  {\it At the same time}, these radiative
corrections automatically induce the dissipative kernels in Eq.(\ref{kernels})
if we use the in-in formalism of quantum field theory.  Note that there is {\it
no ad hoc coarse graining process} at all in this procedure of radiative
corrections.  Actually no information included in the bare Hamiltonian is lost
in the process of radiative corrections.
\item
The system can be {\it consistently expressed in the density matrix formalism
with the ordinary Hilbert space.}  We do not have to extend the original
Hilbert space anymore.  We found that the $V$-particle state spontaneously
decohers.  This is reasonable because the decay of a $V$- particle means not
only a diffusion of energy but also a diffusion of information.
\end{enumerate}
\par
It will be interesting to compare our approach with the others concerning
irreversibility and dissipativity in quantum theory.
\begin{enumerate}
\item
T. Petrosky et al. \cite{petrosky91} and I. E. Antoniou  et al.
\cite{antoniou93}
proposed the extension of the ordinary Hilbert space in order to express the
semi-group property of the evolution of unstable state with hermitian
Hamiltonian.  In our case, instead of extending the representation space, we
extended the field variables making the time integration contour double and
introduced the density matrix in the in-in formalism of quantum field theory.
The generalized Hamiltonian Eq.(\ref{effhamiltonian}) is guaranteed to be
hermitian even if the system is dissipative and irreversible.  A new feature,
which was not discussed in the work \cite{petrosky91} \cite{antoniou93}, is the
destruction of quantum coherence associated with the instability as is
expressed in the last term of Eq.(\ref{efflagrangian}) and
Eq(\ref{effhamiltonian}).  This term directly increases the entropy of the
system as we have seen in Eq.(\ref{lentropy}).\footnote{%
In general, friction term {\it reduces} the entropy and the term which induce
the quantum decoherence (diffusion term) increases the
entropy.\cite{morikawa95b}.
}%
\par
In the work  \cite{petrosky91} \cite{antoniou93}, they tried to connect the
deterministic time-reversible theory and the statistical time-irreversible
theory by a star-unitary operator.  In our case, this kind of transformation is
the process of the generalized radiative corrections in the in-in formalism of
quantum field theory.  According to the work  \cite{petrosky91}
\cite{antoniou93}, a pair of dual spaces was necessary in order to separately
represent the future-decaying and past-decaying states.  In our case, this pair
corresponds to the extension of the variables introducing the doubled time
contour; the time evolution on the forward-time branch represents the
future-decaying state, and vise versa for past-decaying state.
\item
Laplae et al. \cite{laplae65} and Umezawa \cite{umezawa93} introduced a notion
of dynamical map which relates bare fields and the radiatively corrected
asymptotic fields.  This map specifies, among many equivalent representation of
the canonical commutation relation, one representation suitable for the
description of the actual system.  From this point of view, the dynamical map,
in the in-in formalism of quantum field theory, gives an ensemble of equivalent
representations as we see in Eq.(\ref{decomposition}); the total effective
action
$\hat \Gamma[\Phi, \Phi^{\ast}]$, which corresponds to the effective
Hamiltonian Eq.(\ref{effhamiltonian}), is the statistical average of dynamics
each of which has deterministic evolution.  In this sense, our dynamical map
yields one-to-many correspondence instead of one-to-one.
\item
Arimitsu et al. \cite{arimitsu87} derived a general effective Hamiltonian in
the thermo-field dynamics.  Our effective Hamiltonian Eq.(\ref{effhamiltonian})
is similar to that derived by Arimitsu et al. \cite{arimitsu87}.  In their
formalism, it was necessary to double the dynamical degrees of freedom, $\Phi$
fields and $\tilde\Phi$ fields, in order to describe dissipative quantum field
theory within the ordinary Hilbert space.  The situation is almost the same in
our case; we had to introduce $\Phi_+$ as well as $\Phi_-$ fields in order to
describe dissipative quantum field theory within the ordinary Hilbert space.
\item
An usual method to derive quantum-dissipative dynamics is to use the influence
functional method\cite{feynman63}\cite{caldeira83}.  The influence functional
is the induced action by partial trace of an environmental degrees of freedom
and is technically almost the same as our Eq.(\ref{effaction}).  In our case,
we simply considered radiative corrections for all the fields (full trace in
Eq.(\ref{pfunction})) not introducing the environment.  Moreover the
dissipative properties we obtained have nothing to do with the truncation of
the BBGKY hierarchy of greens functions because the Lee model is exactly
solved; one-loop graph is the whole radiative correction.  These facts also
suggest that the dissipative properties are intrinsic to the system.
\end{enumerate}
\par
We summarize our work.  We studied the unstable Lee model constructing the
radiatively corrected effective action Eq.(\ref{effaction}) and Hamiltonian
Eq.(\ref{effhamiltonian}) in the in-in formalism of quantum field theory.
{}From the effective action Eq.(\ref{effaction}), we derived a Langevin
equation for the $V$-field Eq.(\ref{langevin}) which explicitly shows damping
and fluctuation of the state.
{}From the effective Hamiltonian Eq.(\ref{effhamiltonian}), we derived
perpetually increasing entropy Eq.(\ref{lentropy}).
The irreversibility and dissipativity were not manifest in the original bare
Hamiltonian Eq.(\ref{bhamiltonian}).  However we have to make radiative
corrections (dynamical map) which is an indispensable process to define the
asymptotic fields and a feasible theory.  Through this process we obtained the
effective action and Hamiltonian in which the irreversibility and dissipativity
are manifest.  Technically the radiative correction process is regarded as an
averaging process.  This averaging process yields the dissipativity.  However
this averaging process is the unique procedure and does not include any
arbitrariness in principle.  Therefore the dissipativity and irreversibility we
derived are intrinsic to the unstable Lee model.
The increase of entropy is due to the last term in Eq.(\ref{effhamiltonian})
which destroys the quantum coherence (=decoherence) .  This term appears due to
the interference of fields $\Phi_+$ and $\Phi_-$, which is specific to the
in-in formalism of quantum field theory.
In this way irreversibility is consistently described within the ordinary
Hilbert space.  If we force to ascribe a wave function for the unstable state,
then the norm of the wave function would vanish.  We introduced a density
matrix and allowed mixed state for the unstable particle.  Then the probability
is conserved despite the irreversibility.
\par
We would like to report extension of our formalism of intrinsic irreversibility
and dissipativity in our future publications.
\bigskip
% section: acknowledgements
\ni{\bf Acknowledgement}
\par
The author is grateful to Hiroto Kubotani, Izumi Ojima, and Akio Sugamoto for
valuable discussions and  useful comments.  He also would like to thank  The
Kurata Foundation for the Promotion of Science for financial support.
%
%
%
%
%
%
% section: references


\begin{thebibliography}{99}
\bibitem{li86} Ke-Hsueh Li, Phys. Rep. {\bf 134} 1 (1986).
\bibitem{kubotani95} H. Kubotani, T. Okamura, M. Sakagami, Physica A
{\bf 214} 560 (1995).
\bibitem{lee54} T. D. Lee, Phys. Rev. 1329 {\bf 95} (1954).
\bibitem{nakanishi58} N. Nakanishi, Prog. Theor. Phys. {\bf 19}, 607 (1968).
\bibitem{sudarshan78}  E. C. G. Sudarshan, C. B. Chiu, and V. Gorini, Phys.
Rev. {\bf D 18} 2914 (1978).
\bibitem{petrosky91} T. Petrosky, I. Prigogine, and S. Tasaki, Physica {\bf A
173} 175 (1991).
\bibitem{antoniou93} I. E. Antoniou and I. Prigogine, Physica {\bf A 192} 443
(1993).
\bibitem{schwinger61} J. Shwinger, J. Math. Phys. {\bf 2} 407 (1961).
\bibitem{keldysh64} L. V. Keldysh, Sov. Phys. JETP {\bf 20} 1018 (1964).
\bibitem{chou85}  K. Chou, Z. Su, B. Hao, and L. Yu, Phys. Rep. {\bf 118},1
(1985).
\bibitem{morikawa86} M. Morikawa, Phys. Rev. {\bf D33}, 3607 (1986).
\bibitem{morikawa95} M. Morikawa, Prog. Theor. Phys. {\bf 93} 685 (1995).
\bibitem{gamow28} G. A. Gammow, Z. Phys. {\bf 51} 204; {\bf 52} 510 (1928).
\bibitem{morikawa90} M. Morikawa, Phys. Rev. {\bf D42}, 2929 (1990).
\bibitem{morikawa95b}  M. Morikawa, in preparation (1995).
\bibitem{laplae65} L. Laplae, R. N. Sen, and H. Umezawa, Suppl. Prog. Theor.
Phys., Communication issue for the thirty Anniversary of the meson theory by
Dr. H. Yukawa,  637 (1965).
\bibitem{umezawa93} H. Umezawa, {\it Advanced Field Theory --- Micro, Macro,
and Thermal Physics ---},  American Institute of Physics Press  (1993).
\bibitem{arimitsu87} T. Arimitsu and H. Umezawa, Prog. Theor. Phys. {\bf 77} 53
(1987).
\bibitem{feynman63} R. P. Feynman and F. L. Vernon, Ann. Phys. (USA) {\bf 24}
118 (1963)
\bibitem{caldeira83} A. O. Caldeira and A. J. Leggett, Physica {\bf 121A} 587
(1983).
\end{thebibliography}
\end{document}